\documentclass[12pt]{article}
\usepackage{epsfig}

\usepackage{amssymb}
\def\be{\begin{equation}}
\def\ee{\end{equation}}
\def\ba{\begin{array}{c}}
\def\ea{\end{array}}

\def\ben{$$}
\def\een{$$}

\def\bea{\begin{eqnarray}}
\def\eea{\end{eqnarray}}

\def\beax{\begin{eqnarray*}}
\def\eeax{\end{eqnarray*}}

\begin{document}


\vspace{.35cm}

 \begin{center}{\Large \bf

Topology-controlled spectra of imaginary cubic oscillators in the
large$-\ell$ approach

  }\end{center}

\vspace{10mm}

 \begin{center}

 {\bf Miloslav Znojil}

 \vspace{3mm}
Nuclear Physics Institute ASCR,

250 68 \v{R}e\v{z}, Czech Republic

{e-mail: znojil@ujf.cas.cz}


\vspace{13mm}

\end{center}

\vspace{5mm}


 {\bf Abstract}

 \noindent
For quantum (quasi)particles living on complex toboggan-shaped
curves which spread over $N$ Riemann sheets the approximate
evaluation of topology-controlled bound-state energies is shown
feasible. In a cubic-oscillator model the low-lying spectrum is
shown decreasing with winding number $N$.

\newpage

\section{Introduction \label{intr} }

In one of the most up-to-date reviews of the so called ${\cal
PT}-$symmetric quantum mechanics \cite{Roberto} the existence has
been emphasized of multiple connections between certain less usual
differential Schr\"{o}dinger equations for bound states and a number
of important integrable models in statistical physics and/or in
conformal quantum field theory \cite{BLZ}. This correspondence
offers one of explanations of the current growth of interest in many
apparently exotic bound-state problems
 \be
 \left [-\frac{d^2}{dq^2}+V_{eff}(q)
 \right ]\,\varphi_n(q)=E_n\,\varphi_n(q)
 \,
 \label{SEhrovh}
 \ee
where one considers, say~\cite{DDT}, the three-parametric family of
{\em complex} interactions
 \be
 V_{eff}(q) = q^2\,(iq)^{2M-2}-\alpha\,(iq)^{M-1}+\frac{\ell(\ell+1)}{q^2}
 \,,\ \ \ \ \ M>1\,.
  \label{DDTpot}
 \ee
Apparently, this manifestly non-Hermitian model can hardly find any
direct applicability in quantum phenomenology. Fortunately, this
first impression proved wrong. The relevance of similar
non-selfadjoint models has recently been revealed not only in
quantum mechanics (recall, e.g., the so called interacting boson
models in nuclear physics \cite{Geyer}) but also in certain quantum
field theories \cite{Carl} and even in non-quantum nonlinear optics
\cite{optics} (cf. also the other recent applications of
non-Hermitian Hamiltonians collected in review
papers~\cite{Dor,Ali3,SIGMA} and in proceedings \cite{proc}).

Even though the study of similar models is quite common in
mathematics \cite{Sibuya}, virtually all of the above-mentioned
dedicated reviews of their role in physics only appeared very
recently. Indeed, the manifestly non-selfadjoint problems of the
form of Eq.~(\ref{SEhrovh}) + (\ref{DDTpot}) were usually considered
``unphysical". Their studies only appeared in the purely methodical
context of perturbation theory where, typically (cf., e.g.
\cite{BG}), differential Eq.~(\ref{SEhrovh}) + (\ref{DDTpot}) with
$M < 2$ has been assumed integrated along the straight real line or
along its shifted, complexified version
  \be
  q=s-{\rm i}\varepsilon\,\equiv\,q^{(0)}(s)\,,\ \ \ \ \
  s \in (-\infty,\infty)\,.
 \label{line}
  \ee
A decisive progress has been achieved when, in Ref.~\cite{DDT} , the
reality of the spectra generated by potentials (\ref{DDTpot}) has
been given a rather nontrivial proof requiring just the following
very elementary sufficient condition
 \be
 M+1+|2\ell+1|\ > \ \alpha\,.
 \label{subsid}
 \ee
One appreciates, in particular, that for {\em any} coupling $\alpha$
and dominant exponent $2M$ this reality of the energies (i.e., in
principle, their observability) will {\em always} be guaranteed once
we select a sufficiently large strength $\ell\gg 1$ of the
singularity.

The latter observation attracted our attention for two reasons.
Firstly, we kept in mind that the presence of the singularity in the
potential may imply that the individual wave functions
$\varphi_n(q)$ become {\em multisheeted} when treated as
analytically continued functions of complex ``coordinate" $q$
\cite{tobog}. In parallel, we imagined that the condition of
physical acceptability (\ref{subsid}) {\em coincides} with the
condition of mathematical consistency of the so called $1/\ell$
perturbation expansions (for details and further references see
Ref.~\cite{Omar} or Appendix A below).

In what follows we intend to make full use of the latter unexpected
coincidence of the physical and mathematical appeal of the
``strongly spiked", $\ell \gg 1$ versions of potentials
(\ref{DDTpot}). An immediate motivation of our present return to the
related Schr\"{o}dinger Eq.~(\ref{SEhrovh}) arose from
Refs.~\cite{tobog,tobo}. There, the topologically trivial
integration path (\ref{line}) has been replaced by the complex
curves which were allowed to encircle the singularities of the
potentials (cf., e.g., Refs.~\cite{584} and \cite{identifi} or Sec.
\ref{4.} below for a compact introduction to this new possibility of
quantum model-building).

Unfortunately, even in the simplest models~(\ref{SEhrovh}) the
tobogganic bound-state problems were found extraordinarily hard to
solve numerically \cite{Wessels,Bila}. For this reason, no
sufficiently reliable illustration of a tobogganic spectrum is at
our disposal at present. At the same time, the above-mentioned tests
of applicability of $1/\ell$ perturbation expansions to
non-tobogganic non-Hermitian models~\cite{Omar} may be perceived as
a source of new optimism and as a decisive encouragement of our
present project of an approximative evaluation of the tobogganic
low-lying spectra.

Our results will solely involve tobogganic versions of the concrete
toy model (\ref{DDTpot}) possessing the single singularity in the
origin $q=0$ and considered just at a fixed sample exponent $M=3/2$
and at the simplest coupling $\alpha=0$. Under these assumptions our
main attention will be paid to our Schr\"{o}dinger
Eq.~(\ref{SEhrovh}) integrated along the whole family of nontrivial,
``tobogganic" curves denoted by the symbol $q^{(N)}(s)$ and
classified, in an exhaustive manner, by their winding number
$N=0,1,2,\ldots$ (cf. their sample in Figure~\ref{firsone}). By
definition (cf. Sections \ref{4.} and \ref{4a} below) these smooth
complex curves will interconnect an $N-$plet of sheets of the
Riemann surface ${\cal R}$ supporting the analytic, multivalued wave
functions~$\varphi_n(q)$.

\begin{figure}[h]                     
\begin{center}                         
\epsfig{file=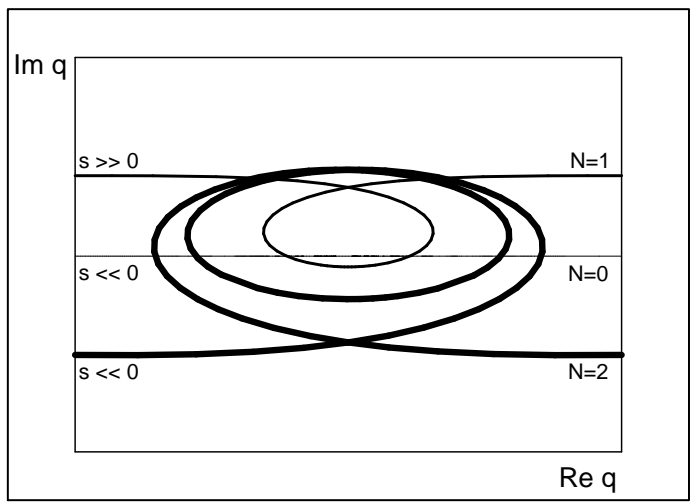,angle=270,width=0.6\textwidth}
\end{center}                         
\vspace{-2mm} \caption{The complex straight contour $q^{(0)}(s)$ of
eq.~(\ref{line}) (thin line) and its first two winding descendants
$q^{(1)}(s)$ (medium curve) and $q^{(2)}(s)$ (thick curve).
 \label{firsone}}
\end{figure}

We shall reflect the dominance of the centrifugal repulsion near the
origin as an encouragement of the use of the approximative
large$-\ell$ method of the solution of Eq.~(\ref{SEhrovh}). The
essence of this method will be explained in Section \ref{5.}. For
our $M = 3/2$ (i.e., imaginary cubic) quantum toboggan the
application of this method will finally be described in Section
\ref{6.} complemented by a few comments and a brief summary in
Section~\ref{7.}.

\section{Topologically nontrivial integration paths\label{4.}}

Several important theoretical reasons of the contemporary growing
interest in non-tobogganic as well as tobogganic models sampled by
Eqs.~(\ref{SEhrovh}) + (\ref{DDTpot}) + (\ref{line}) deserve to be
re-emphasized. One of the most typical ones lies in the emergence of
several unexpected conceptual links between the ODE language and its
correspondence to the physical theories of integrable statistical
and field-theoretical models. In this sense, in a way mentioned in
\cite{Roberto,DDT}, the integrable SUSY point of the sine-Gordon
model has been found related to the value of the exponent $M=1/2$ in
(\ref{DDTpot}). Similarly, the partnership with the Yang–Lee model
emerged at $M=3/2$, etc (cf. section 5.6 in \cite{Dor} for a ``full
dictionary").

One should also mention the importance of the flow of inspiration in
the opposite direction. Indeed, certain very specific questions
asked in the context of conformal field theory \cite{BLZ} brought,
as their direct consequence, the immediate turn of attention from
the most elementary and regular $\ell=0$ versions of Schr\"{o}dinger
Eq.~(\ref{DDTpot}) to its specific  singular generalizations. A
concrete example  of the emerging family of relevant generic
potentials with multiple centrifugal-type singularities may be
found, e.g., as Eq. (1) in Ref.~\cite{BLZ}.

Naturally, after similar generalizations and incorporation of
singularities even the ODE mathematics becomes perceivably less
transparent (cf., e.g., the classification problem solved for the
first nontrivial double-singularity tobogganic model in \cite{584}).
Indeed, under the presence of singularities in the potential
$V_{eff}(q)$ the wave functions $\varphi_n(q)$ generated by the
corresponding Schr\"{o}dinger Eq.~(\ref{SEhrovh}) get analytically
continued to a topologically nontrivial, punctured and {\em
multisheeted} Riemann surface ${\cal R}$. Even the description of
this surface itself (characterized by its so called monodromy group
\cite{Zelondek}) strongly depends on the details of the form of the
potential.

In the majority of the above-mentioned applications, the technical
obstacle emerging with nontrivial monodromies is most often
circumvented via an {\em ad hoc} constraint which guarantees a
``trivialization" of the topological structure of ${\cal R}$ and of
its monodromy \cite{patr}. An explicit form of the ``trivialization"
constraint is sampled, e.g., by Eq. (3) in Ref.~\cite{BLZ} or by
Eq.~(6.31) in Ref.~\cite{Dor}. In this context, new an interesting
questions emerge, of course, when one {\em admits} a nontrivial
monodromy~\cite{tobog,tobo}.

The latter step of generalization is now in the center of our
interest. In general, it would induce a  rich menu of curves
supporting the wavefunctions and connecting, in general, different
sheets of the surface ${\cal R}$ \cite{584}. In the spirit of an
older letter \cite{BT}, {\em different} spectrum may be attributed
to {\em each} attachment of a new tobogganic curve to the same
differential Schr\"{o}dinger equation. In the two-singularity case,
for illustration, the menu of these curves has been made available
via a specific interactive demonstration using MATHEMATICA by
Novotn\'{y} \cite{Novotny}.

In this context we feel curious what happens to the spectrum after
the changes of topology in the most elementary though still
nontrivial single-singularity case.

\section{Paths encircling single singularity\label{4a}}

For the sake of definiteness we may define our curves of
integration, say, by formula Nr. (10) of Ref.~\cite{identifi},
 \be
 q(s)=q^{(N)}(s)=-{\rm i}\left [{\rm i}(s-{\rm i}\varepsilon)
 \right ]^{2N+1}\,,\ \ \ \ \
  s \in (-\infty,\infty)\,.
  \label{kurvicka}
 \ee
Without any change of the spectrum these curves may be also smoothly
deformed of course (cf. Figures 5 or 6 in Ref. \cite{Roberto} for
illustration). In this case we only have to preserve both the left
and right asymptotic parts of the curve $q^{(N)}(s)$ unchanged (a
detailed discussion of this point may be also found in
\cite{Roberto}).

From the point of view of physics (discussed in more detail, say, in
\cite{SIGMA,identifi}) one of the most important features of the
complexified quantum systems represented by Eqs. (\ref{SEhrovh}) +
(\ref{DDTpot}) on a line (\ref{line}) (with $\varepsilon \neq 0$) or
on a curve (\ref{kurvicka}) can be seen in the manifest loss of
observability of their ``coordinate" $q \notin \mathbb{R}$. In
principle, the only measurable quantity remains to be the
bound-state energy spectrum. This clarifies the main purpose of
considering our specific class of a non-Hermitian models at
nonvanishing winding numbers $N=1,2,\ldots$: We may expect that the
new free parameter $N$ opens a new, potential-independent way of the
modification of the spectrum needed, say, during its better fit to
some given experimental data (of course, just hypothetical data in
our present methodical study).

In this broad framework the purpose of our present paper lies in
finding a sample of {\em quantitative} analysis of the
$N-$dependence of the energies. The two main sources of inspiration
of such a study may be seen in the existence of a gap between the
older, non-tobogganic results of Refs.~\cite{BBjmp} [reproduced also
in Figure 8 of review \cite{Roberto} and using, incidentally, just
$\ell=\alpha=0$ in Eq.(\ref{DDTpot})] and the B\'{\i}la's
preliminary tobogganic numerical results which were presented in his
very recent paper~\cite{Bila} and in his dissertation \cite{disse}.

On the technical level the main ingredient of our present
calculations will lie in the rectification transformation which will
replace our imaginary cubic differential Schr\"{o}dinger equation
defined along one of the winding trajectories (\ref{kurvicka}) by
another, equivalent differential Schr\"{o}dinger-type equation
obtained via a suitable change of variables.

In a more explicit specification of such a transformation we shall

\begin{itemize}

\item
distinguish between the non-winding curves where $N=0$ [abbreviating
$q^{(0)}(s)\,\equiv\,y(s)$ in Eq.~(\ref{line})] and the genuine
winding-paths  $q^{(N)}(s)$ with $N=1,2,\ldots$ [use
Eq.~(\ref{kurvicka}) and abbreviate $q^{(N)}(s)\,\equiv\,z(s)$ at
any $N \geq 1$ ],

\item
start from the initial cubic and tobogganic bound-state problem
 \be
 \left [-\frac{d^2}{dz^2}
 +\frac{\ell(\ell+1)}{z^2}+{\rm
 i}z^3\right ]\,\varphi^{[N]}_n(z)=E^{[N]}_n\,\varphi^{[N]}_n(z)
 \,,
 \label{SErrrhbe}
 \ee

\item
employ the change of variables $z=-{\rm i}\,({\rm i}y)^{2N+1}$ or,
in opposite direction, ${\rm i}y = ({\rm i}z)^\alpha$ where
$\alpha={1}/({2N+1})$ is rational, i.e., where several (logarithmic)
Riemann sheets of the original variable $z$ are mapped into the
single (cut) complex plane of the new variable $y$.

\end{itemize}

 \noindent
The detailed realization of such a project may be summarized as the
direct use of the elementary substitution rules
 \be
  dz=(-1)^N\,({2N+1}) \,y^{2N}\,dy\,,\ \ \ \ \ \ \
 \frac{d}{dz}=
 \frac{(-1)^N}{(2N+1)\,y^{2N}}\,
 \frac{d}{dy}
 \label{chovbcd}
 \ee
 \be
  \varphi_n^{[N]}(z)=
  y^{N}\,\psi_n(y)\,,
 \ \ \ \ \ \ \
 L=({2N+1})\,
 \left (\ell+
 \frac{1}{2}
 \right )-\frac{1}{2}
 \,
 \ee
under which our original ordinary differential Eq.~(\ref{SErrrhbe})
becomes transformed into the following Sturm-Schr\"{o}dinger
\cite{shend} differential equation
 \be
 \left [-\frac{d^2}{dy^2}
 +\frac{L(L+1)}{y^2}
 +{\rm
 i}\,(-1)^N\,(2N+1)^2\,y^{10 N+3}
 \right ]\,\psi_n(y)
 = (2N+1)^2\,
 y^{4N}\,E^{[N]}_n\,\psi_n(y)
 \,.
 \label{SErov}
 \ee
This change of variables realizes the one-to-one correspondence not
only between the two respective (in general, logarithmic) Riemann
surfaces of $z$ and $y$ but also between the original integration
path (\ref{kurvicka}) (living on the original, tobogganic
$z-$surface) and the new integration path for $y$ which, by its
present specification, coincides with the straight-line formula
(\ref{line}).

During the search for the energies $E^{[N]}_n$ the new,
``rectified", generalized eigenvalue problem (\ref{SErov}) +
(\ref{line}) must still be addressed by purely numerical means in
general. Still, the key merit of its choice is that one can employ
some more or less standard methods. In particular, we intend to use
here an appropriately adapted form of the large$-\ell$ approximation
method of Ref.~\cite{Omar}, to be first briefly summarized in
Sec.~\ref{5.} and subsequently applied to our problem in
Sec.~\ref{6.}.

\section{The large$-\ell$ approximation technique\label{5.}}

The key ingredients of the approximations where the quantity
$1/\ell$ remains small may be most easily illustrated via the
non-tobogganic, single-sheeted version of our imaginary cubic
oscillator, integrated along the straight-line path $ q =
q^{(0)}(s)$ of Eq.~(\ref{line}),
 \be
 \left [-\frac{d^2}{dq^2}
 +\frac{\ell(\ell+1)}{q^2}+{\rm
 i}q^3\right ]\,\varphi^{[0]}_n(q)=E^{[0]}_n\,\varphi^{[0]}_n(q)
 \,.
 \label{SErhbe}
 \ee
The main idea of the approximative search for eigenvalues
$E^{[0]}_n$ is based on the invariance of the spectrum during any
parallel shift of the path $q^{(0)}(s)$ \cite{Roberto}. Thus, we are
allowed to choose the free parameter $\varepsilon=\varepsilon_j$ in
Eq.~(\ref{line}) in such a way that one and only one of the roots
$q=Q=Q_j$, $j=1,2,\ldots$ of the auxiliary equation $\partial _Q
V_{eff}(Q)=0$ will lie on the corresponding specific, $j-$dependent
exceptional line of our ``coordinates" in Eq.~(\ref{SErhbe}),
$q^{(0)}(s) \to q^{(0)}_j(s) $.

In the next step we may rewrite our auxiliary equation in its fully
explicit form $\ 2 \ell(\ell+1)=3{\rm i}Q^5$ and notice that in the
vicinity of {\em any} of the quintuplet of its roots  $Q=Q_j$, $j =
1,2,\ldots,5$ we may Taylor-expand our effective potential in terms
of the new, shifted complex variable $\xi = q-Q$,
 \be
  V_{eff}(q) =    V_{eff}(Q) +  \frac{1}{2}\,V_{eff}''(q)\,\xi^2
     +  \frac{1}{6}\,V_{eff}'''(q)\,\xi^3+
   \ldots
    \,.
    \label{tay}
  \ee
Next, we have to take into account the available explicit formulae
for our $V_{eff}$ and for all the roots
 \be
    Q_j= -{\rm i}\,\tau\,\exp(\frac{2{\rm i}\pi (j-1)}{5})\,,
    \ \ \ \ \ \tau=
      \left |
  \left (
    {\frac{2}{3}\,{\ell(\ell+1)}}
    \right )^{1/5}\,
 \right |
 \,, \ \ \ \ \ j = 1,2,\ldots,5\,.
 \ee
Due to our assumption of the smallness of the quantities $1/\ell$
and/or $1/\tau$ we can choose $j=1$ and simplify our Taylor
series~(\ref{tay}) to its much more explicit form
 \be
  V_{eff}(q) =   -\frac{5}{2}\,\tau^3 + \frac{15 \,\tau}{2}\,\xi^2
   - 5\,{\rm i}\,\xi^3  +
    {\cal O}(\tau^{-1})
    \,.
    \label{tayl}
  \ee
This formula shows that up to the negligible error term our
potential coincides with the deep and real harmonic-oscillator well,
complemented just by a purely imaginary cubic perturbation.

In the spirit of perturbation theory we shall accept our tentative
choice of $j=1$ in what follows. The more general discussion of
difficulties emerging in connection with the use of the other roots
with $j>1$ may be found in Refs.~\cite{Omar} and \cite{dd}. As long
as we now have the new independent variable $\xi = q^{(0)}_1(s)-Q_1\
\equiv\ s$ which is real, we may insert the Taylor series
(\ref{tayl}) in our Schr\"{o}dinger Eq.~(\ref{SErhbe}) and rewrite
it in its leading-order anharmonic-oscillator reincarnation
 \be
 \left [-\frac{d^2}{d\xi^2} -\frac{5}{2}\,\tau^3 + \frac{15 \,\tau}{2}\,\xi^2
   - 5\,{\rm i}\,\xi^3  +
    {\cal O}(\tau^{-1})
 \right ]\,\varphi^{[0]}(-{\rm i}\tau + \xi)
 =
 E_n^{[0]}
 \,\varphi^{[0]}(-{\rm i}\tau + \xi)
 \,.
 \label{SEaphrovhbe}
 \ee
In the final step of our considerations we introduce another very
small auxiliary quantity $\sigma = 1/\tau^{1/4}$ and rescale the
coordinate $\xi \to \sigma \xi$. This yields the final version of
our approximate eigenvalue problem,
 \be
 \left [-\frac{d^2}{d\xi^2} -\frac{5}{2\,\sigma^{10}}
 + \frac{15}{2}\,\xi^2
   - 5\,\sigma^5 \,{\rm i}\,\xi^3  +
    {\cal O}(\sigma^6)
 \right ]\,\varphi^{[0]}(-{\rm i}\tau + \xi)
 =
 \sigma^2 \,E_n^{[0]}
 \,\varphi^{[0]}(-{\rm i}\tau + \xi)
 \,.
 \label{Shrovh}
 \ee
As long as $\sigma \ll 1$ is very small,  the leading-order version
of this equation has the purely harmonic-oscillator form
 \be
 \left [-\frac{d^2}{d\xi^2} -\frac{5}{2\,\sigma^{10}}
 + \frac{15}{2}\,\xi^2
    +
    {\cal O}(\sigma^5)
 \right ]\,\varphi^{[0]}(-{\rm i}\tau + \xi)
 =
 \sigma^2 \,E_n^{[0]}
 \,\varphi^{[0]}(-{\rm i}\tau + \xi)
 \,
 \label{Shrovhdom}
 \ee
which is exactly solvable.

As long as we are interested here just in the very global features
of our sample of tobogganic spectrum (like, e.g., the leading-order
winding-number-dependence of the removal of its degeneracy, etc), we
shall skip here the construction of higher-order corrections
completely. Interested readers may find an instructive sample of
such calculations in Ref.~\cite{Omardva}. In this case we may
already conclude that the low-lying spectrum evaluates to the
following formula
 \ben
 E_n^{[0]}
  =-\frac{5\,\tau^{3}}{2}
  +\sqrt{\frac{15\,\tau}{2}}\,(2n+1)
 +
    {\cal O}(\tau^{-3/4})\,,\ \ \ n = 1, 2, \ldots\,
 \een
with an asymptotically vanishing error term and predicting the
perceivably negative and approximately equidistant leading-order
spectrum.


\section{The large$-\ell$ approximants at tobogganic $N>0$\label{6.}}


On the basis of our preceding non-tobogganic $N=0$ results we are
now prepared to move to the genuine tobogganic models with windings
$N=1,2,\ldots$ entering our rectified Sturm-Schr\"{o}dinger
differential Eq.~(\ref{SErov}). Immediately, we may make use of the
fact that one of the most convenient small parameters $\rho =
1/(\ell+1/2)^2$ remains $N-$independent. This allows us to keep the
$N-$dependence carried  solely by the adapted shifts
$\varepsilon=\varepsilon^{(N)}$ in the straight lines $ y (s)$ of
Eq.~(\ref{line}).

In the first step we have to Taylor-expand our effective potential
again,
 \ben
 V_{eff}[y(x))]= \frac{L(L+1)}{y^2}
 +{\rm
 i}\,(-1)^N\,(2N+1)^2\,y^{10 N+3}\,.
  \een
In contrast to section \ref{5.} an amended notation must be
introduced now since the original tobogganic (i.e., multisheeted)
$z-$reference complex points $Q_j= q^{(N)}_j(0)$ differ from their
transformed, Sturm-Schr\"{o}dinger maps or partners $T_j=
q^{(0)}_j(0)$. The latter points lie on the rectified, straight-line
contours (\ref{line}) in the simpler complex $y-$plane with,
possibly, just a single cut directed upwards.

The positions of the latter reference points will again be
determined by the derivative-vanishing condition
 $
  \partial_y V_{eff}(y)=0
 $.
Thus, for our particular dynamical model each $L-$dependent root
$y=T=T_j(L)$ of this equation will coincide with one of the $10N+5$
distinct roots of the following elementary algebraic equation
 \ben
 2L(L+1)=(2N+1)^2(10N+3)\,T^{10N+5}\,.
 \een
All of these roots form the vertices of a regular $(10N+5)-$angle in
complex plane. At $N=0$ we would just return to the pentagon which
has been obtained  in the non-tobogganic imaginary cubic model
above. At any other $N\geq 0$ we shall abbreviate
  \be
     \tau=\tau^{(N)}=
      \left |
  \left (
    \frac{2\,L\,(L+1)}
    {(2N+1)^2(10N+3)}
    \right )^{1/(10N+5)}\,
 \right |
 \,
 \label{pantau}
 \ee
and set
 \ben
 T_1=-{\rm i}\tau\,, \ \ \ \
 T_{j}=T_{j-1}\,e^{2{\rm i} \pi/(10N+5)  }\,,\ \ \
 j=2, 3,
 \ldots, 10N+5\,.
 \een
In a way which parallels our above discussion of the model with
$N=0$ we use just the same exceptional root $T_1=-{\rm i}\tau$
giving the maximal size of the shift $\varepsilon = \tau$ in
Eq.~(\ref{line}). This choice also offers the single isolated {\em
real} minimum of our effective potential along this line so that we
may work again with the adapted real independent variable $\xi=s$
everywhere in what follows.


After a somewhat tedious calculation one obtains the Taylor-series
asymptotic approximation of $V_{eff}(y)$ at any $N$,
  \ben
  V_{eff}(y) =  -\frac{1}{2}\,(2N+1)^2(10N+5)\,\tau^{10N+3} +
  \omega^2_{(N)}\,\tau^{10N+1}\,\xi^2
     + \mu(N)\,\tau^{10N}\,\xi^3+
    {\cal O}(\tau^{10N-1})
    \,
  \een
with
 \ben
    \omega_{(N)}=(2N+1)\,\cdot
    \sqrt{\frac{(10N+3)(10N+5)}
    {2}
    }\,.
    \een
Similar closed formulae can also be derived for $\mu(N)$, etc. This
means that our rectified Schr\"{o}dinger equation will acquire the
following approximate $N-$dependent asymptotic form
 \be
 \left [-\frac{d^2}{d\xi^2}
    +
  \omega^2_{(N)}\,\tau^{10N+1}\,\xi^2  + \mu(N)\,\tau^{10N}\,\xi^3+\ldots
  \right ]\,\psi_n(-{\rm i}\tau + \xi) =
 \label{SErovna}
  \ee
  \ben
  = \left \{
  E^{[N]}_n\,(2N+1)^2\,
 \left[ \tau^{4N}
 +\ldots\right ]
  +\frac{1}{2}\,(2N+1)^2(10N+5)\,\tau^{10N+3}
 \right \}
 \,\psi_n(-{\rm i}\tau + \xi)
 \,.
 \een
In a preliminary step of analysis we notice that the
anharmonic-oscillator corrections proportional to the second
coupling $\mu(N)$ will remain asymptotically negligible and that
they may be ignored. Thus, in the asymptotic regime where $\tau \gg
1$ we shall apply the same considerations as tested in Appendix~A.

\begin{figure}[h]                     
\begin{center}                         
\epsfig{file=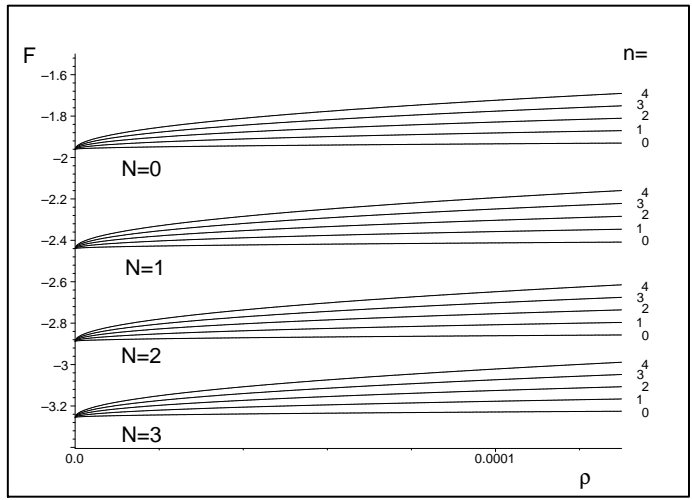,angle=270,width=0.6\textwidth}
\end{center}                         
\vspace{-2mm} \caption{ Tobogganic spectra (\ref{main}) at $\rho =
1/(\ell+1/2)^2$. The first five lowest levels are rescaled $\left [
F_n^{(N)}=\rho^{3/5}\,E_n^{(N)},\,n=0,1,2,3,4\right ]$ and displayed
at the first four winding numbers $N=0,1,2,3$.
 \label{fionse}}
\end{figure}

In an explicit error analysis we rescale $\xi \to \sigma \xi$ where
$\sigma = \tau^{-(10N+1)/4}$. This leads to the equation
 \be
 \left [-\frac{d^2}{d\xi^2} -\frac{5}{2\,\sigma^{10}}
 + \frac{15}{2}\,\xi^2
   - 5\,\sigma^5 \,{\rm i}\,\xi^3  +
    {\cal O}(\sigma^6)
 \right ]\,\varphi^{[N]}(-{\rm i}\tau + \xi)
 =
 \sigma^2 \,E_n^{[N]}\,\varphi^{[N]}(-{\rm i}\tau + \xi)
 \,
 \label{Shrovhbe}
 \ee
which shows that although
the asymptotic order of magnitude of the energies
remains $n-$ and $N-$independent,
$E_n^{[N]} \propto {\cal O}(\ell^{6/5})$,
there is in fact no degeneracy in $N$ and that also the
degeneracy in $n$ becomes removed by the next-order correction
(cf. illustrative Figure \ref{fionse}).

\begin{figure}[h]                     
\begin{center}                         
\epsfig{file=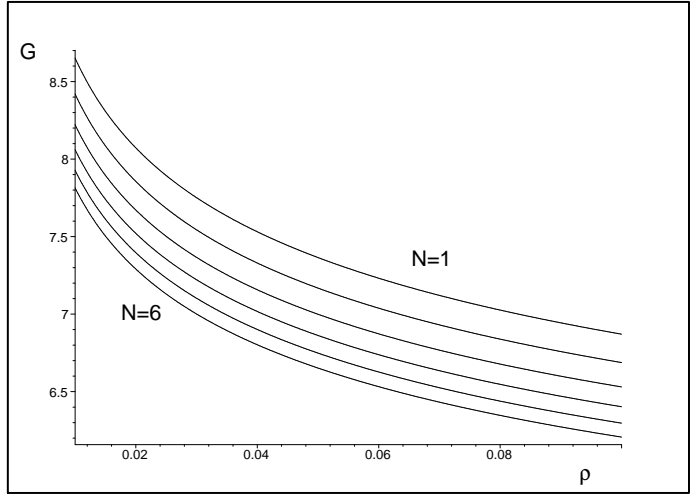,angle=270,width=0.6\textwidth}
\end{center}                         
\vspace{-2mm} \caption{The decrease of the level spacings
$E_{n+1}^{[N]}- E_n^{[N]}\ \equiv\ G^{(N)}$.
 \label{druha}}
\end{figure}

In a more explicit asymptotic study of our model we have to keep in
mind that our large auxiliary parameter $\tau$ also changes with the
winding number $N$, $\tau=\tau_{(N)}$. The source of this dependence
does not lie only in the manifest presence of $N$ in
formula~(\ref{pantau}) but also in the $N-$dependence of the
upper-case parameter $L$. This is a mere abbreviation of the
expression $L=L(N,\ell)=({2N+1})(\ell+1/2) -{1}/{2}$ in which one
recognizes the presence of the small variable $\rho =
1/(\ell+1/2)^2$ used in  Figures \ref{fionse} or \ref{druha}. Thus,
the physics hidden behind the asymptotics in $L$ can be re-read as a
combined centrifugal-repulsion effect where both the coupling
strength $1/\rho$ and the number of rotations $N$ appear to act in a
coherent manner.

At a fixed topological characteristic $N$ of the tobogganic quantum
system in question our asymptotically simplified equations indicate
that the level-splitting will predominantly be mediated by the
harmonic-oscillator term $\xi^2$ in Eq.~(\ref{Shrovhbe}). Thus, the
leading-order equidistance of the levels (cf. Figure \ref{druha})
will  only be broken, in the next-order approximation, by the cubic
term ${\cal O}(\xi^3)$. An explicit estimate of the size of the
latter correction is still feasible. From Eq.~(\ref{Shrovhbe}) one
immediately deduces that the  anharmonic-oscillator part of our
potential will contribute by the corrections of the order ${\cal
O}(\tau^{-(10N+5)/4})$. They may be perceived, say, as a
higher-order modification of the coefficient $\omega_{(N)}$.

We should add that another specific feature of the tobogganic
problem emerges
in connection with the nontriviality of
the
right-hand-side ``weight" term
 \ben
 W(y)=(2N+1)^2\,
 y^{4N}\,
 \een
entering Eq.~(\ref{SErov}) at $N\geq 1$. Fortunately, the scaling argument
can be re-applied here to give the estimate of the
first non-constant $\xi-$dependent correction to $W(T)$ in the form
${\cal O}(\tau^{-(14 N+7)/4}))$. This correction
is even smaller than the contribution of the anharmonic force
so that it can either be neglected (on the present level of precision)
or, whenever needed, calculated and incorporated, say, by the appropriately
adapted
systematic perturbation techniques as described in
Ref.~\cite{Omar}.

We may now return to Eq.~(\ref{SErovna}) and summarize our observations
in the closed formula for the energies,
 \ben
 E_n^{[N]}
  =-\frac{10N+5}{2} \,\tau^{6N+3}
  +
  \ \ \ \ \ \ \ \ \ \
  \ \ \ \ \ \ \ \ \ \
  \ \ \ \ \ \ \ \ \ \
  \ \ \ \ \ \ \ \ \ \
  \ \ \ \ \ \ \ \ \ \
  \een
   \be
  +\frac{2n+1}{2N+1}\,\cdot
    \sqrt{\frac{(10N+3)(10N+5)}
    {2}
    }\,\tau^{N+1/2}
 +
    {\cal O}(\tau^{-(6N+3)/4})\,
    \label{main}
 \ee
giving
the  low-lying spectrum at $n = 1, 2, \ldots\,$. This is our
final result. Once we recollect that
$\tau \propto \ell^{2/(10N+5)}$ this formula reconfirms
our above estimate that $ E_n^{[N]}\propto \ell^{6/5}$, i.e., that
in the dominant order
the power-law exponent of $\ell$ does not vary with $N$.
In this setting the formula specifies
the numerical factor
and clarifies the mechanism of the removal
of the degeneracy between topologically
nonequivalent toboggans.

In the subdominant, next-order level of precision
the next numerical coefficient in the spectral formula
will still retain an explicit form.
The dominant part of the distance $ G^{(N)}=E_{n+1}^{[N]}-
E_n^{[N]}$ between levels appears to have the power-law
form with another constant exponent, $ G^{(N)}\propto
\ell^{1/5}$. This is well illustrated by Figure \ref{druha} where we
see how the differences between constants $\lim_{\ell \to
\infty}\, G^{(N)}/\ell^{1/5}$
remove the topology-independence of the gaps, i.e., their leading-order
degeneracy in $N$.

This effect slightly weakens with the growth of winding number
$N\geq 1$. It is worth adding that the function $
G^{(N)}/\ell^{1/5}$ of $N$ happens to have its maximum at $N=1/2$ so
that the observed monotonicity in $N$ does not apply to the
``anomalous", non-winding curve $G^{(0)}/\ell^{1/5}$ which,
incidentally, lies somewhere  in between $G^{(2)}/\ell^{1/5}$ and
$G^{(3)}/\ell^{1/5}$.


\section{Summary \label{7.} }

Our present paper offers the first quantitative confirmation of the
hypothesis that the spectrum of quantum toboggans can be well
controlled not only by the potential and, independently, by the
suitably complexified asymptotic boundary conditions but, in
addition, also by some suitable topological, non-asymptotic
characteristics of the curves of the complexified ``coordinates".
These characteristics were shown to lead to the potentially useful
changes in the spectra.

For the sake of brevity we analyzed only the simplest nontrivial
quantum toboggans which are characterized by the presence of a
single, isolated branch point singularity of $\psi(q)$ in the
origin. In this case the non-equivalent toboggans can be classified
by the mere integer winding number $N$. Of course, this does not
mean that the analysis of the spectrum is easy. For illustration we
may mention the recent diploma work \cite{Wessels} where, for
several potentials, the stability and applicability of a few most
common numerical methods has been shown restricted just to the
non-winding dynamical regime with $N=0$.

More stable though still preliminary numerical results are available
in Ref.~\cite{Bila} where the first encouraging samples of the real
spectra of toboggans were presented using $N \leq 2$. Although this
analysis remained restricted just to the smallest exponents $M\in
(1,3)$, an empirical confirmation of the reality of the energies at
$M=2$ has been achieved. This seems to be in correspondence with
Ref.~\cite{Shindva} where the exceptional role of integer exponents
$M$ in certain non-tobogganic models has been revealed.

Intuitively one could expect that in the tobogganic cases the domain
of a robust reality of the energies will move to the higher
exponents $M \geq M^{(N)}$. Indeed, at $N=0$, the explicit numerical
confirmation of the rightward shift of the boundary of the interval
of the acceptable  exponents can be found in Refs.~\cite{BBjmp}.
Unfortunately, our present approach  is unsuitable for an analysis
of similar phenomena since, e.g., Eq.~(\ref{subsid}) controls the
reality of the energies just via the insensitive sum of a given $M$
with our very large $|2\ell+1|$.

On the side of the advantages of our present large$-\ell$ recipe we
may mention the transparency of the determination of the unique and
optimal line (\ref{line}) of complex coordinates. In the leading
order the use of this path reduces the dominant part of the
effective interactions $V_{eff}(q) $ to the solvable harmonic
oscillator. In addition, whenever needed, the series of subdominant
corrections can be evaluated in recurrent manner. In this sense the
message of our present paper is non-numerical, showing that the
low-lying levels of our toboggans can be given by closed formula and
that they appear to vary sufficiently strongly with the variation of
the topological winding number $N$.

\subsection*{Acknowledgements}

Supported by the Institutional Research Plan AV0Z10480505, by the
M\v{S}MT ``Doppler Institute" project LC06002 and by GA\v{C}R grant
Nr. 202/07/1307.

\newpage

\section*{Appendix A. The large$-\ell$ approximation in an exactly solvable model}

Let us consider, in a test of the method, the exactly solvable
example of the linear harmonic oscillator
 \be
 V_{eff}^{(HO)}[q(s))]= \frac{\ell(\ell+1)}{(s-{\rm i}\varepsilon)^2}+\omega^2\,
 (s-{\rm i}\varepsilon)^2
 \label{toy}
  \ee
in the strongly spiked dynamical regime, $\ell \gg 1$. In a
preparatory step of the analysis at large $\ell$ let us
Taylor-expand this effective potential near its (complex) stationary
point $q=Q^{(HO)}$ such that
 \ben
  \partial_q V_{eff}^{(HO)}(q)=0\,.
 \een
For our model every such a point is determined by the elementary
algebraic equation $\ \ell(\ell+1)=\omega^2T^4$ with an easy
solution,
  \be
    Q=Q_j= (-{\rm i})^j\,\tau
 \,,\ \ \  j = 1,2,3,4\,,\ \ \ \ \tau=
  \left |\sqrt{\frac{1}{\omega}\sqrt{\ell(\ell+1)}}\,
 \right |\,.
 \label{square}
 \ee
All of these four candidates for the (in general, complex) point
where $V_{eff}(q)$ attains its (local) minimum also satisfy the
sufficient condition since
 the real part of the second derivative
 of the effective potential
 remains $Q-$independent and
positive at all $q=Q$ since
 $ V''_{eff}(Q)=8\omega^2$.

In the next step let us define the distances $\xi = q-Q$ from the
minima and generate the Taylor series
 \ben
  V_{eff}(q) =  V_{eff}(Q) + \frac{1}{2}  V''_{eff}(Q) \,\xi^2
  + \frac{1}{6}  V'''_{eff}(Q)\,\xi^3 +
    \ldots
       \,.
  \een
For our toy model this gives
 \ben
  V_{eff}(q) =   2\omega^2Q^2 + 4\omega^2\,\xi^2
  + \frac{4}{Q}\omega^2\,\xi^3  +
    {\cal O}(Q^{-2})
    \,.
  \een
Under the assumption that $\ell$ is large, i.e.,  $|Q|\gg 1$, the
insertion of this truncated series would transform our  bound-state
problem (\ref{SEhrovh}) into its approximatively isospectral
partner.

The unrestricted freedom of the choice of the parameter
$\varepsilon>0$ in  Eq.~(\ref{line}) enables us to set
$\varepsilon=\tau$ and to let the line $q^{(0)}(s)$ cross the point
$Q_1= -{\rm i}\,\tau$. This is equivalent to the replacement of our
initial Schr\"{o}dinger equation (\ref{SEhrovh}), at all the
sufficiently large $|\ell|\gg 1 $ and/or $\tau\gg 1$, by the new
equation
 \be
 \left [-\frac{d^2}{d\xi^2}-2\omega^2\,\tau^2
 + 4\omega^2\,\xi^2    +
    {\cal O}(\tau^{-1})
 \right ]\,\varphi_n(-{\rm i}\tau + \xi)
 =
 E_n(\omega)
 \,\varphi_n(-{\rm i}\tau + \xi)
 \,.
 \label{SEaphrovha}
 \ee
In the leading-order approximation this replacement leads to the new
closed formula
 \ben
 E_n(\omega)
 =
 -\omega
 \sqrt{(2\ell+1)^2-1}
 + 2\omega\,(2n+1)+
    {\cal O}(\ell^{-2/3})
 \,,\ \ \ n = 1, 2, \ldots\,.
 \een
It is rather amazing to see how well this $\ell\gg 1$ estimate
reproduces the {\em exact} low-lying spectrum
 \ben
 E_n(\omega)=\omega\,\left (
 4n+1- 2\ell
 \right )\,,\ \ \ n = 1, 2, \ldots\,, n_{max}\,, \ \ \
 n_{max}< \ell+1/2
 \een
as derived, for our toy model (\ref{toy}), in Ref.~\cite{ptho}.

\end{document}